\documentclass{jps-cp}

\title{Dark matter, dark photons through the observables}

\author{Gennady \textsc{Kozlov} }

\inst{Joint Institute for Nuclear Research, Joliot-Curie st. 6, Dubna, 141980, Russia }

\email{kozlov@jinr.ru}

\date{March 02, 2020}

\abst{We analyse  dark matter (DM) produced somehow in space-time is charged under  the hidden $U^{\prime} (1)$ gauge symmetry and interacting with Standard Model (SM) through
 the scale invariance breaking sector containing dilatons and dark photons (DP).   We find the solutions for DM and DP in terms of observables. The DM observable is under DP field shape influence. DP is the natural mediator between DM and the SM sectors. The phenomenology of DP physics and the registration of dark photons are discussed. }

\kword{Dark matter, dark photon, dilaton.}

\begin{document}
\maketitle

\section{Introduction}

The study of dark matter (DM) in direct or non direct searches is actual and ongoing. Dark photon (DP) as one of the fundamental vector forms of DM in sub-GeV region becomes the main target to this activity (see, e.g., [1], [2] and the refs. therein). DP may have its origin in spontaneous breaking of conformal symmetry  [3]:  dark photons may have been produced by decays of dilatons, the scalar fields related to breaking scale invariance. If dilatons are typically abundant in Nature, their dynamics is expected to provide the generic initial conditions for production of dark photons after breaking the conformal symmetry. On the other hand, there is the possibility that dilaton fields themselves constitute the observed DP abundance. Here, we have to be careful, because in the absence of interactions the dilaton field may carry the features of dipole "ghost" with an indefinite metric. In this case, the Fourier transformation of two-point Wightman function (TPWF) for dilaton fields contains $\theta (p^{0})\delta^{\prime} (p^{2})$ distribution  [3] in the momentum $p$ - space $S(\Re_{4})$, and TPWF does not have the sense in terms of the positive measure. The latter also comes from the dilatation transformations to TPWF. In addition, the presence of $ \delta^{\prime} (p^{2})$ is a consequence of the non-unitarity of the translations relevant to dilaton.  To avoid the unphysical result we use the $S$-matrix approach to asymptotic states of dilaton mixed with the Standard Model (SM) Higgs boson (Sec. 3).

One of the basic theoretical instruments to search  both DM and DP is the extension of SM within the principles of conformal (scale) and gauge invariances. The dilaton acquires mass from explicit breaking of scale invariance in the scale-invariant sector. The mass of the dilaton is small compared to the energy scale relevant to spontaneous breaking the scale invariance.  DM can be in the states  with various possible spins: spin-1/2, scalar and as a gauge boson. Here, we consider the spin 1/2 DM that is treated as a Dirac particle. We study the model where both DM and DP reside in a hidden sector where the latter also contains a dilaton field. The key question is to find the signature of DM and  DP in terms of observables.

The plan of the paper is as follows. The Sec. 2 is devoted to breaking of symmetry. The mixing between the Higgs boson and the dilaton associated with the dipole "ghost" is considered in Sec 3. In Sec. 4 the model with one derivative scalar is developed. In Sec. 5 we consider the model of the dilaton containing conformal  anomaly. In Sec. 6 the Abelian Dipole Dark Photon Model (ADDPM) with different couplings is presented. The Sec. 7 is devoted to DM and DP in terms of observables where the latter are the only under stochastic forces influence (Sec. 8). In Sec. 9 some comments to phenomenology of dark photons are given. The paper is concluded by Sec. 10.  The results are given in the forms that are convenient to author, and clarify the main aspects which will not require to refer to original papers. 

\section{Breaking of symmetry}
It is known the gap between the conformal sector and the SM sector in the sense of matter fields in terms of observables. 
We propose dynamical evolution of dilaton and DP masses in the early Universe that choose a scale parametrically smaller than the cutoff of the theory. We assume the dilaton mass and DP mass  equal to zero are the  special points in terms of dynamics but not in terms of symmetry. There are the points where the weak forces spontaneously break the symmetry and theory enters another stage of interaction between DM and SM.  
%It is known the gap between the conformal sector and the SM sector in the sense of matter fields in terms of observables. 

In the approximate conformal theory, the derivative of dilatation current $S_{\mu}(x)$ is almost equal to zero, $\partial_{\mu} S^{\mu} (x) = \theta^{\mu}_{\mu} (x) \simeq 0$, where $S^{\mu} =\theta^{\mu\nu}\,x_{\nu}$ under the dilatation transformation $x_{\mu}\rightarrow \omega\,x_{\mu}$, $\omega$ is an arbitrary constant; $\theta_{\mu\nu}$ is the energy-momentum tensor, and $\theta^{\mu}_{\mu}$ is its trace. In SM, $\theta_{\mu}^{\mu} \neq 0$ flashing the breaking of scale invariance due to running couplings and dimensional transmutation.
In papers [4]-[6] there were shown that  vector field with non zero mass occurred in case of special conformal transformations (SCT) does not describe the independent vacuum fluctuations, and, thus, this vector field is an excessive (not necessary) field in the theory. In paper [7] the inverse Higgs condition has been suggested, that can allow to express the excessive field in terms of "physical" one where the latter can already describe the independent fluctuations of the vacuum. This mechanism does admit to replace the vector field by some composition of the dilaton field within SCT. 

In field theories, conformal symmetry is the maximal space-time group symmetry (supersymmetry does not considering here). The scheme of spontaneous symmetry breaking is 
\begin{equation}
\label{e1}
G_{con} (d) \times G_{int} \rightarrow \mho,
\end{equation}
where $G_{con}(d)$  is the conformal group of dimension $d$; $G_{int}$ is the group of internal symmetry and $\mho$ contains the vector subgroups of both space-time and internal symmetries. 
To understand the relation between DP and the dilaton, one needs  the theory based on an effective Lagrangian density (LD) governed by spontaneous breaking of both conformal invariance and gauge invariance. One can choose the field order parameter $\Phi$ the vacuum average of which breaks the conformal invariance at some scale $f$.  In some cases $\Phi$  may be nothing other but the dilaton field. The factor-space relevant to (\ref{e1}) has the form 
%\begin{equation}
%\label{e2}
$$ g_{con} = e^{i\,P_{\mu}x^{\mu}}\,e^{i\,K_{\nu}B^{\nu}(x)}\,e^{i\,D\,\Phi (x)} , $$
%\end{equation}
where $P_{\mu}$,  $K_{\mu}$ and $D$ are generators of translations, SCT and dilatations, respectively; $x_{\mu}$ are coordinates in the space considered; $B_{\nu} (x)$ is
%and $\phi (x)$ are 
the  vector field that may play the special role: it is the DP field in spontaneous breaking phase governed by the effective theory. 
The origin of DP is the conformal anomaly, and an evidence of DP is through the decay of a dilaton. A light narrow resonance is therefore one of features of nearly conformal dynamics with production of two dark photons.
The operators $\hat K_{\mu}$ and $\hat D$ acting on the order parameter $\Phi$ give
\begin{equation}
\label{e3}
\hat K_{\mu}\Phi = 2\,x_{\mu}\,d_{\Phi}\,\Phi,\,\,\,\, \hat D \Phi = d_{\Phi}\, \Phi  ,
\end{equation}
where $d_{\Phi}$ is the scale dimension of $\Phi$ if the latter is the space-homogeneous field. From (\ref{e3}) one can no see the direct action of SCT on $\Phi$, just through the space-time coordinates $x_{\mu}$. It means the vector field $B_{\mu} (x)$ itself may not be necessary to describe any local fluctuations of the vacuum. According to the formalism of inverse Higgs mechanism [7], $B_{\nu} (x) \sim - f^{-1}\partial_{\nu} \Phi (x)$, where the vector field is nothing other but the derivative of the dilaton field. We shall come back to this result in  Sec.   6 where ADDPM [3]  is developed. 
%In the next section, we consider on the classical level the model of interaction the spin 1/2 and vector fields with the scalar (dilaton-like) field through its derivative.
%\begin{equation}
%\label{e4}
%B_{\mu}(x) = \frac{g}{m^{2}}\,J_{\mu} (x) - \frac{1}{m}\,\partial_{\mu} \varphi (x) + \frac{\eta}{m^{3}}\,\partial_{\mu} \Box\varphi (x),
%\end{equation} 
%where $g$ is the coupling between DP and Dark Matter (DM) spinor field in the covariant derivative $D_{\mu} = \partial_{\mu} + i\,g\,B_{\mu}$, $J_{\mu}( x)$ is the current of DM field, $%\varphi (x)$ is the dilaton field and $m$ is the mass of DP. So, the inverse Higgs condition is the second term in (\ref{e4}) for DP field in ADDPM. The solution (\ref{e4}) indicates the %correct number of degrees of freedom in the effective theory. 
 
 \section{Higgs-dilaton mixing and dipole "ghost"}
 In the conformal sector the  spectrum may contain the mass eigenstates of the Higgs boson and the dilaton through their mixing which depends on the ultraviolet completion of the theory. A distinctive difference between the Higgs boson and the dilaton is in their couplings to massless gauge bosons. If there are two Higgs doublets, one Higgs is responsible to heavy masses of third generation quarks, another one is responsible to the masses of light quarks, and the dilaton mixes with just the former Higgs. We can admit the dilaton field may carry the properties of dipole "ghost" field [3] defined in the space with indefinite metric. Thus, we deal with general transform of the physical state in a mixture with physical and "ghost' states. To avoid the unphysical result we use the $S$-matrix approach to the theory and consider the local scalar field $X(x)$ in the form 
$$X (x) = z(x) + \sum_{n} c_{n}\,\zeta_{n} (x), \,\,\, c_{n} = const,$$
where $z (x)$ is the set of physical states including the Higgs boson;
% the dilaton field $\varphi (x)$ may stand as a virtual (fictitious) state.
$\zeta_{n}(x)\in \{\zeta_{1} (x), ..., \Phi (x), ...\}$  and some of the fields in $\zeta_{n}$ may have a negative probability; the dilaton field $\Phi (x)$ may stand as a virtual (fictitious) state. The total Hilbert space  $\mathcal {H} (\Re^{4})$  is the sum of the Hilbert space  $ S (\Re^{4})$ of the physical states $z(x)$ and of that one $S^{\prime} (\Re^{4})$ for $\zeta_{n}(x)$ states.
  
 For the Higgs-dilaton mixing let us introduce the perturbation (distortion) to the Higgs boson field $H(x)$ by
 \begin{equation}
\label{e33}
X(x) = \frac{1}{\sqrt{2}}\,H(x) + i\,\kappa\,\Phi(x),\,\,\,\,\,\, X^{\star}(x) = \frac{1}{\sqrt{2}}\,H(x) - i\,\kappa\,\Phi(x)  ,
\end{equation}
where $\kappa $ is the distortion strength. The expressions (\ref{e33}) are nothing other but the fluctuations by the dilaton field around the vacuum expectation value of the Higgs boson $\langle H\rangle = 246\, GeV$. 
We can write down LD containing two scalar fields $X$ and $X^{\star}$ with complex conjugate masses $\mu$ and $\mu^{\star}$ and with interaction part given by DP field $B_{\mu}$:
 \begin{equation}
\label{e333}
L = \frac{1}{2\,i} \left (\partial_{\mu}X\partial^{\mu} X - \mu^{2}X^{2} \right ) - \frac{1}{2\,i} \left (\partial_{\mu}X^{\star}\partial^{\mu} X^{\star} - \mu^{2^{\star}}X^{2^{\star}} \right ) + D_{\mu} XD^{\mu} X^{\star}  ,
\end{equation}
where $D_{\mu} = \partial_{\mu} + igB_{\mu}$. Having in mind (\ref{e33}),  LD (\ref{e333}) becomes
% \begin{equation}
%\label{e31}
%L = L_{s} + L_{int},
%\end{equation}
%where 
 \begin{equation}
\label{e32}
 L= \frac{1}{2}\partial_{\mu}H\partial^{\mu} H + \kappa\,\partial_{\mu}H\partial^{\mu}\Phi - \kappa\,\mu^{2}_{1}\,H\Phi - \frac{1}{2}\mu^{2}_{2}\, H^{2} +
     g\kappa B_{\mu} (\partial^{\mu}\Phi\, H - \partial^{\mu} H\,\Phi ) + \frac{1}{2} g^{2} B^{2}_{\mu} H^{2} .
\end{equation}
 %\begin{equation}
%\label{e321}
%L_{int} =  g\kappa B_{\mu} (\partial^{\mu}\Phi\, H - \partial^{\mu} H\,\Phi ) + \frac{1}{2} g^{2} B^{2}_{\mu} H^{2}. 
%\end{equation}
Here, $\mu^{2}_{1} = \Re e\mu^{2}$, $\mu^{2}_{2} = \Im m\mu^{2}$, and the terms $\sim \kappa^{2} << 1 $ (small fluctuations) are neglected in (\ref{e32}). The model with LD (\ref{e32}) is not invariant under transformations
$$\Phi (x)\rightarrow\Phi (x) + \Lambda(x), \,\,  H (x)\rightarrow H (x) + \Lambda(x),\,\, B_{\mu} (x)\rightarrow B_{\mu} (x) + \partial_{\mu}\Lambda (x)$$
unless $\Lambda (x) = 0$. The equations of motion are 
%\begin{equation}
%\label{e331}
$$ \left [ \Box + \mu^{2}_{1} - g (\partial B) - 2g B_{\mu}\partial^{\mu} \right ] \Phi = -\kappa^{-1} (\Box + \mu^{2}_{2} - g^{2} B^{2}_{\mu} ) H, $$
%\end{equation}
%\begin{equation}
%\label{e341}
$$ \left [ \Box + \mu^{2}_{1} + g (\partial B) + 2g B_{\mu}\partial^{\mu} \right ] H = 0. $$
%\end{equation}
If one neglects the interaction by $B_{\mu}$ field the canonical formalism implies the commutator $[H(x),H(x^{\prime})] = 0$, and
the solution for the dilaton field $\Phi$ is given through the Higgs boson field $H$:
 \begin{equation}
\label{e351}
\Phi (x) = \frac{1}{\kappa} \left (\frac{\mu^{2}_{2}}{\mu^{2}_{1}} \right ) \left ( 1 + \frac{1}{2} x_{\mu}\partial^{\mu} \right ) H(x) + C(x),
\end{equation}
where the free (auxiliary) field $C(x)$ obeys Eq. $(\Box + \mu^{2}_{1} ) C(x) = 0.$ It is easy to check that the solution of the dilaton field  is given by the Fourier transform
$$\Phi (x) = \int \left [ \Phi (k) + \frac{1}{\kappa} \left (\frac{\mu^{2}_{2}}{\mu^{2}_{1}}  \right ) \left ( 1 - \frac{1}{2} ikx\right )  H (k) \right ] e^{-ikx}\, \delta (k^{2} - \mu^{2}_{1})\, d_{4} k .$$
%The following commutators can be found easily:
%$$ [H(x), \Phi (x^{\prime})] = i\,\Delta (x-x^{\prime}; \mu^{2}_{1}),\,\,\,  [\Phi(x), \Phi (x^{\prime})] = \frac{i}{\kappa} \left (\frac{\mu^{2}_{2}}{\mu^{2}_{1}} \right ) \left ( 1 + \frac{1}{2} x_{\mu}%\partial^{\mu} \right ) \Delta (x-x^{\prime}; \mu^{2}_{1}), $$
%where $\Delta (x, \mu^{2}_{1}) = 2\pi i\int d_{4} p\, sgn (p^{0})\,\delta (p^{2} - \mu^{2}_{1})\, e^{-i p x}$. 

Actually, $\Phi $ can be interpreted in terms of dipole "ghost" field in case of very weak couplings to DP 
 \begin{equation}
\label{e361}
{\left (\Box + \mu^{2}_{1} \right )}^{2}\Phi (x) \simeq 0.
\end{equation}
The origin of the dipole "ghost" behaviour of the dilaton field is the second term in  (\ref{e351}) $\sim x_{\mu}\partial^{\mu} H(x)$.
In this case, $\Phi (x)$ is part of some asymptotic pattern of a more complicated theory containing other fields corresponding to the physical metric. 
%The total Hilbert space is the direct %product of the Hilbert space of the "ghost" fields (with an indefinite metric) and of that one of the physical fields with usual metric. 
The properties of $\Phi (x)$ in (\ref{e361}) may be found in [8]. The massless case of the model with (\ref{e361}) is more singular.
 It is easy to see that the dilaton field $\Phi$ in (\ref{e351}) is the Higgs boson field in classical sense distorted by "ghost" Higgs. 

%We suppose that each amplitude of the state possesses by both physical and virtual parts, however the part of the amplitude corresponding to $\zeta$ state is defined unique by its %physical part based on the state $z$. 
We'll use the $S$-matrix approach to the operator form of $X(x)$ (see (\ref{e33})):
$$\hat X = \hat H + \kappa\,\hat \Phi, $$
where  $\hat H = P\,\hat X $ ($\hat H \in S(\Re ^{4})$) and $\kappa\,\hat\Phi = (1 - P)\,\hat X$ ($\hat\Phi \in S^{\prime}(\Re ^{4})$). The operator $P$  projects the states $\hat H$ from $\mathcal {H} (\Re^{4})$ to $S (\Re ^{4})$; $P ^{+} = P$, $ P^2 = P$; ${\parallel\hat X\parallel}^2 = {\parallel \hat H \parallel}^2 + \kappa^{2}{\parallel\hat\Phi\parallel}^2$, ${\parallel \hat H \parallel}^2 > 0$. 
In the $S$-matrix approach $\hat X_{+ \infty} = S \,\hat X_{- \infty}$. 
%In $S$-matrix approach  $(XX^{\star})_{\infty} = z_{-\infty} + \zeta_{-\infty}$ at $t\rightarrow - \infty$ first, and because of interactions, the system turns into the state $\Phi_{+\infty} = %z_{+\infty} + \zeta_{+\infty}$ at $t\rightarrow + \infty$. The following condition $\parallel\Phi_{-\infty}\parallel = \parallel\Phi_{+\infty}\parallel$ is evident.
% Then one finds 
%\begin{equation}
%\label{e44}
%z_{+ \infty} = P\,S (z_{- \infty} + \zeta_{- \infty}),
%\end{equation}
%\begin{equation}
%\label{e45}
%\{ \zeta_{- \infty} + (1 - P)\,S (z_{- \infty} + \zeta _{-\infty})\} = 0,
%\end{equation}
%where the nonlocal boundary condition $\zeta _{- \infty} + e^{i\,\delta}\,\zeta _{+ \infty} = 0$ is used ($\delta$ is a phase).  Eq. (\ref{e45}) defines the unphysical (virtual) part of the %amplitude at $t\rightarrow -\infty$ by the physical part of the amplitude
%$$\zeta_{-\infty} = - \{1 + (1 - P)\,S\}^{-1}\, (1 - P)\,S\, z_{-\infty}.$$
%The input physical states of the system is described by the vector states of the form:
%$$\Phi_{-\infty} = z_{-\infty}  - \{1 + (1 - P)\,S\}^{-1}\, (1 - P)\,S\, z_{-\infty}.$$
%From (\ref{e44}) and (\ref{e45}) 
As the result, we find the asymptotic "ghost" dipole state  $\kappa\,\hat \Phi _{+\infty}$ via the asymptotic state $\hat H_{-\infty}$ of the (physical) Higgs boson
%\begin{equation}
%\label{e46}
 $$\kappa\,\hat\Phi_{+ \infty} = \{ 1 + (1 - P)\,S \}^{-1} (1- P)\,S\, \hat H_{- \infty}, $$
%\end{equation}
where  $\hat H_{- \infty}$ is defined from the equation $\hat H_{+\infty} = \tilde S\, \hat H_{-\infty}$ and the unitary matrix $\tilde S$ is 
$\tilde S = PS \{ 1 + (1 - P) S\}^{-1}$. The latter connects the physical amplitude of the Higgs state operator $\hat H$ only.
As to the application to light flavour sector beyond SM, the dilatons could perhaps be the origin of the observed KOTO excess in $K_{L}\rightarrow \pi^{0}\nu\bar\nu$ decay [9]. The minimal Higgs portal mixed with dilatons can explain this anomaly. This is due to extra decays  $K_{L}\rightarrow \pi^{0} X$, where $X$ is long-lived and weakly interacting scalar so that it appears as missing energy at KOTO [10].

 \section{One derivative scalar model}
 We shall not quantise DM and DP fields in the conventional manner for two reasons:\\
 - to be consistent with experiment where no such particles have been identified,\\
 - to avoid the infra-red (IR) problems in perturbation theory.\\
 DP field can be removed from LD in favour of interaction with DM.  Consider the classical variant of one derivative scalar LD
 \begin{equation}
\label{e4}
L = \bar\chi (i\partial_{\mu}\gamma^{\mu} - m_{\chi} - g B_{\mu}\gamma^{\mu} )\chi + \partial_{\mu}\tilde\varphi\,\partial^{\mu} b + \frac{1}{2} b^{2} - I^{\mu} (B_{\mu} - \partial_{\mu}\tilde\varphi),
\end{equation}
 where $\chi$ is a Dirac fermion singlet (with the mass $m_{\chi}$) charged under $U^{\prime} (1)$ group, a candidate to DM; $B_{\mu}$ is the $U^{\prime} (1)$ gauge field, a candidate to DP and $g$ is its coupling to DM; $\tilde\varphi$ is the sub-canonical massless scalar field; the field $b$  plays the role of a gauge-fixing multiplier and it remains free; $I_{\mu}$ is an auxiliary arbitrary vector field. The Eq. of motion $\partial_{\mu}\tilde\varphi (x) = B_{\mu}( x)$ transforms (\ref{e4}) to 
% \begin{equation}
%\label{e5}
$$ L = \bar\chi (i\partial_{\mu}\gamma^{\mu} - m_{\chi} )\chi  - g\,J_{\mu}\, \partial^{\mu}\tilde\varphi + \partial_{\mu}\tilde\varphi\,\partial^{\mu} b + \frac{1}{2} b^{2}, $$
%\end{equation}
 where $J_{\mu} = \bar\chi\gamma_{\mu}\chi$. One can easily find the equations of motion for $\chi$ and $\tilde\varphi$:
 \begin{equation}
\label{e6}
(i\partial_{\mu}\gamma^{\mu} - m_{\chi} )\chi = g\gamma^{\mu}\,\partial_{\mu}\tilde\varphi\cdot\chi,
\end{equation}
 \begin{equation}
\label{e7}
\Box^{2}\tilde\varphi (x) = 0.
\end{equation}
In quantum case the solution of (\ref{e6}) is 
\begin{equation}
\label{e8}
\chi (x) = :e^{-i\,g\,\tilde\varphi (x)}:\chi^{(0)} (x),
\end{equation}
where $\chi^{(0)} (x)$ is the solution of free Dirac equation. The normal product $: ...:$ in (\ref{e8}) stands for 
%\begin{equation}
%\label{e9}
$$ :e^{-i\,g\,\tilde\varphi (x)}: = e^{-i\,g\,\tilde\varphi^{(+)}(x)} e^{-i\,g\,\tilde\varphi^{(-)} (x)}, $$
%\end{equation}
where $\tilde\varphi {(x)}$ is decomposed into negative-frequency (annihilation) and positive-frequency (creation) parts
%\begin{equation}
%\label{e10}
$$\tilde\varphi (x) = \tilde\varphi^{(+)}(x) + \tilde\varphi^{(-)} (x),\,\,\,\, \tilde\varphi^{(+)}(x) = {\left [ \tilde\varphi^{(-)} (x)\right ]}^{+}. $$
%\end{equation}
Here $\tilde\varphi^{(-)} (x)\Omega = 0$, $\langle\Omega\vert\Omega\rangle = 1$ for the vacuum state $\Omega$. 
The equations of motion $b = \Box\tilde\varphi$ and $\Box b = 0$ accompanied by the following equal-time commutators of the fields $b(x)$ and $\tilde\varphi (x)$ 
$$ {[\partial_{0} b(x), \tilde\varphi (y) ]}_{\vert_{x^{0} = y^{0}}} = {[\partial_{0} \tilde\varphi(x), b (y) ]}_{\vert_{x^{0} = y^{0}}} = \frac{1}{i} \delta (\vec x - \vec y)$$
lead to the basic commutator of the dilaton field
\begin{equation}
\label{e101}
[\tilde\varphi (x), \tilde\varphi (y)] = 2\pi\int sgn (p^{0})\,\delta^{\prime} (p^{2})\,e^{-i\,p(x-y)}\,d_{4} p = \frac{1}{8\pi i}\, \theta (z^{2})\,sgn (z^{0}),\,\,\, z = x-y,
\end{equation}
where the distribution $sgn (p^{0})\,\delta^{\prime} (p^{2})$ in (\ref{e101}) is well-defined in terms of odd homogeneous generalised function in Minkovsky space $S^{\prime} (\mathbb{M})$ of temperate generalised functions. There is a well-known singularity $\delta^{\prime}(p^{2})$ in (\ref{e101}) corresponding to dipole field $\tilde\varphi (x)$ satisfying Eq. (\ref{e7}). The breaking of  gauge symmetry requires such a type of singularity in TPWF of vector potential in Abelian Higgs model [11]. The TPWF $W(x)  = \langle\Omega\vert \tilde\varphi (x)\,\tilde\varphi (0)\vert \Omega\rangle $ is
%of $\tilde\varphi (x) $ is 
%The commutator $[\tilde\varphi^{(-)} (x), \tilde\varphi^{(+)} (y)]$ is the two-point Wightman function $W (x-y) = \langle\Omega\vert \tilde\varphi (x)\,\tilde\varphi (y)\vert \Omega\rangle$ %by commuting $\tilde\varphi^{(-)} (x)$ to the right and $\tilde\varphi^{(+)} (x)$ to the left: 
\begin{equation}
\label{e111}
W(x)  =  2\pi\int \theta (p^{0})\delta^{\prime} (p^{2})e^{-ipx} d_{4}p = \frac{-1}{(4\pi)^{2}} \left [\ln \vert k^{2} x_{\mu}^{2}\vert + i\pi\epsilon (x^{0})\theta (x^{2})\right ].
\end{equation}
Here, $k$ is a positive constant with dimension of mass, the parameter of IR regularisation. The commutator $[\tilde\varphi^{(-)} (x), \tilde\varphi^{(+)} (0)]$ is $W (x)$ by commuting $\tilde\varphi^{(-)} (x)$ to the right and $\tilde\varphi^{(+)} (x)$ to the left.
All the other commutators are  $[\tilde\varphi^{(-)} (x), \tilde\varphi^{(-)} (y)] = [\tilde\varphi^{(+)} (x), \tilde\varphi^{(+)} (y)] = 0$. 

The generalised function (distribution)  $\theta (p^{0})\,\delta^{\prime} (p^{2})$ in (\ref{e111}) is defined only on the basic functions $D(p)$ that are equal to zero at $ p = 0$.
We suppose $D(p)$ is the test function related to dilaton field $\Phi (p)$ with the mass $\mu$, and $D(p = 0) = 0$. Than, one can write down 
\begin{equation}
\label{e1111}
2\pi \int \theta (p^{0})\delta^{\prime} (p^{2})\, D (p) d_{4}p = -\frac{\partial}{\partial\mu^{2}}  2\pi \int \theta (p^{0})\delta (p^{2} - \mu^{2})\, D (p) d_{4}p = -\frac{\partial}{\partial\mu^{2}}  \int  D (p) \frac{d_{3}\vec p}{2\,p^{0}}. 
\end{equation}
The result of (\ref{e1111}) is 
$$2\pi \int \theta (p^{0})\delta^{\prime} (p^{2})\, D (p) d_{4}p = \int_{\Gamma_{0}^{+}} \frac{1}{2 n\,p} \left  (- n\partial_{p} + \frac{1}{n\,p}\right ) D (p) \frac{d_{3}\vec p}{2\,p^{0}}, $$
where $n$ is the fixed vector from $V^{+} = \{p \in \mathbb{M},\, p^{2} > 0, \,p^{0} > 0 \}$  with the property $n^{2} = 1$, $\Gamma_{\mu}^{+} = \{p \in \mathbb{M}, \,p^{2} = \mu^{2}, \,p^{0} > 0 \}$, $\mu > 0$.  

The Fourier transformation of (\ref{e111}) itself does not have a positive measure. It also follows from (\ref{e111}) under the dilatation transformation 
$$ W(x) \rightarrow W(\omega x) = W (x) - \frac{1}{2(2\pi)^{2}}\,\ln\omega,\,\,\,\, \omega > 0.$$
Thus, the representation of $\tilde\varphi (x)$ obeying (\ref{e7}) has to be formulated in the space with indefinite metric using the pseudo-Hilbert space. 
%In [2] it has been shown the indefinite metric emerged in Green's functions does not appear in asymptotic expressions for physical observables. 

 \section{Dilaton with conformal anomaly}
 In this section, we investigate the approximately scale invariant sector with the effects stemming from conformal anomaly that involves SM, scalar dilaton, DM and DP, and reflects the violation of conformal invariance of hidden sector. Both, DM and DP are lightest species of some hidden sector with its own gauge interactions and some symmetry that stabilises DM and DP. Consider the LD of the effective theory with a cutoff $\sim 4\pi f$:
 \begin{equation}
\label{e11}
L_{1} = \sum_{l:q,\chi} \bar l (i\hat D_{l} - m_{l}) l - b(\partial  B) + \frac{1}{2} b^{2} - I^{\mu} (B_{\mu} + e^{-\sigma}\,\partial_{\mu}\sigma),
\end{equation}
where spin 1/2 fields $l$ run over quarks $q$ and DM $\chi$ with masses $m_{l}$; $\hat D_{l} = D^{\mu}_{l}\gamma_{\mu} = (\partial_{\mu} + i\,g_{l}B_{\mu})\gamma^{\mu}$, $g_{l}$ are couplings of $q$ and $\chi$ with DP field $B_{\mu}$; $I_{\mu}$ is an auxiliary vector field.  The DM field $\chi$ is neutral under SM quantum numbers and charged under a hidden gauge symmetry that is broken at low energies. The dilaton field $\sigma (x)$ has non-linearly transformation $\sigma (x) \rightarrow\sigma (x\,e^{\omega}) + \omega$ under dilatation transformation of coordinates $x_{\mu} \rightarrow \omega\,x_{\mu}$. To avoid the non-linear term $e^{-\sigma}\,\partial_{\mu}\sigma$ in (\ref{e11}) one can make the redefinition $\sigma (x) \rightarrow \Phi (x) = f\,e^{-\sigma (x)}$. 

LD (\ref{e11}) is invariant under restricted gauge transformations:
\begin{equation}
\label{e12}
l\rightarrow l\, e^{-i\,g_{l}\,\Lambda}\,\,\, (l: q,\chi),\,\,   B_{\mu}\rightarrow B_{\mu} + \partial_{\mu}\Lambda,\,\,\,
\tilde\varphi \rightarrow \tilde\varphi + \Lambda, \,\,\,\, b\rightarrow b,\,\,\,\, I_{\mu}\rightarrow I_{\mu},
\end{equation}
where    $\tilde\varphi (x) = \Phi (x)/f$,   $\Lambda (x)$ satisfies $\Box\Lambda (x) = 0$. The only Dirac particles $l$ lighter than dilaton are included in (\ref{e11}) because of conformal condition for the first coefficients of $\beta$-function (for details, see [12], [13]). If the quantum system is close to phase transition or to the critical point, $\Phi (x)$ spontaneously slipped from having zero average value in the hot space domain to the one having an average value $f$. The commutators of fields involved in LD ({\ref{e11}) and the relevant two-point Wightman functions, the Green's functions can be easily obtained using the lessons given in the paper [14].

DM has no direct couplings to SM sector. The dilaton is the dominant messenger between SM and DM, however, DM thermal relic abundance is governed  by its coupling to the derivative of dilaton field [3], and thus, strictly related to $f$. The dilaton has the coupling to the trace $\theta^{\mu}_{\mu} (x)$ of the energy-momentum tensor containing all the fields in the scale-invariant sector, and may also pick up the couplings at the loop level due to scale anomaly. The dilaton may  be lighter than DM, and dark photons be the dominant decay product of a dilaton.
At energies below $4\pi f$ we have the following couplings of a dilaton to SM and a hidden sector
\begin{equation}
\label{e13}
L_{2} = \tilde\varphi \left (\theta^{\mu}_{\mu_{tree}} + \theta^{\mu}_{{\mu}_{anom}}\right ),
\end{equation}
where 
\begin{equation}
\label{e141}
\theta^{\mu}_{\mu_{tree}}  = - \sum_{light\,\, l:q,\chi} m_{l}\,\bar l\,l  .
\end{equation}
The contribution from massive gauge bosons are neglected in (\ref{e141}). The couplings of $\tilde\varphi$ to massless gauge bosons (photons, gluons) and massive DP are generated at loop level and are given by the anomaly term to (\ref{e13})
\begin{equation}
\label{e15}
\theta^{\mu}_{{\mu}_{anom}} = - \frac{\alpha}{8\pi}\, b_{EM}\, F_{\mu\nu}F^{\mu\nu} - \frac{\alpha_{s}}{8\pi} \, b_{0}\,G^{a}_{\mu\nu}G^{\mu\nu\,a} - \frac{\bar\epsilon}{8\pi} F_{\mu\nu}B^{\mu\nu}.
\end{equation}
Here, $F_{\mu\nu} = \partial_{\mu}A_{\nu} - \partial_{\nu}A_{\mu}$ with the field $A_{\mu}$ of the photon, $B_{\mu\nu} = \partial_{\mu}B_{\nu} - \partial_{\nu}B_{\mu}$ with the field $B_{\mu}$ of DP;  $\alpha$ and $\alpha_{s}$ are the fine and strong coupling constants, respectively; $b_{EM}$ and $b_{0}$ are the coefficients of electromagnetic (EM) and strong (QCD) $\beta$-functions (to be defined below);   $\bar\epsilon\sim O(\varepsilon)$, $\varepsilon$  is free parameter, kinetic mixing angle between the photon and DP. 
Thus, LD of the model  $L = L_{1} + L_{2}$ is invariant under (\ref{e12}) in addition to $A_{\mu}\rightarrow A_{\mu} + \partial_{\mu}\Lambda$.
%$$l\rightarrow l\, e^{-i\,g_{l}\,\Lambda},\,\,\, (l: q,\chi),\,\, A_{\mu}\rightarrow A_{\mu} + \partial_{\mu}\Lambda,\,\,  B_{\mu}\rightarrow B_{\mu} + \partial_{\mu}\Lambda,
%\tilde\varphi \rightarrow \tilde\varphi + \Lambda, \,\, b\rightarrow b,\,\, I_{\mu}\rightarrow I_{\mu},$$
%where  $\Lambda (x)$ satisfies $\Box\Lambda (x) = 0$. 

If SM is embedded in conformal sector, the total $\beta$-functions for EM and strong interactions vanish above $4\pi f$, so, $b_{EM}$ and $b_{0}$ are computed from contributions of particles lighter than the dilaton. In particular,  $b_{EM} = - 80/9 $ if the dilaton mass is less than two masses of the W-boson [15] ;  $b_{0}=  -11 + 2 n_{light} / 3$, where $n_{light}$ is the number of quarks lighter than the dilaton. We assume the contribution from DP is very small since $\varepsilon\sim 10^{-3} - 10^{-7}$ (see, e.g., [1], [3] and the refs. therein). 
Using the equations of motion relevant to LD $L$, one can find the equation for the physical dilaton field through the trace anomaly and kinetic mixing terms:
\begin{equation}
\label{e161}
\theta^{\mu}_{{\mu}_{tree}}  + \theta^{\mu}_{{\mu}_{anom}} = \left ( \Box^{2} + \frac{\bar\epsilon}{4\pi}\, F_{\nu\mu}\,\partial^{\mu}\partial^{\nu}\right )\tilde\varphi.
\end{equation}
%If the conformal symmetry is restored, the dilaton $\tilde\varphi$ becomes the "ghost" canonical field obeying the equation $\Box^{2}\tilde\varphi (x) = 0$. 
The nature of the scale anomaly is seeing in the chiral limit in (\ref{e161})
%\begin{equation}
%\label{e17}
$$ \theta^{\mu}_{{\mu}_{anom}} =  \Box (\partial\cdot B) + \frac{\bar\epsilon}{4\pi}\, F_{\nu\mu}\,\partial^{\mu} B^{\nu} $$
%\end{equation}
as the expansion over derivatives of DP field $B_{\mu} (x)$. If the conformal symmetry is restored, the dilaton $\tilde\varphi$ becomes the "ghost" canonical field obeying the dipole equation $\Box^{2}\tilde\varphi (x) = 0$. We propose, the detectable messenger between DM and SM is a scalar dilaton through its decay to dark photons. DP in its solution is a composition of derivatives of the dilaton [3] (see the next section). 
 
\section{Abelian Dipole Dark Photon Model with different couplings}
 
In this section we proceed in brief to ADDPM [3] where it is  assumed the approximate scale invariance of ADDPM for energies below $\sim 4\pi f$ and where the scale invariance is preserved by the dilaton field $\Phi$. 
%The dilaton acquires mass from breaking of the symmetry. The mass of the dilaton is small compared to the energy scale $f$. 
Consider  LD 
\begin{equation}
\label{e181}
L = -\frac{1}{4} F_{\mu\nu}^{2} - \frac{\varepsilon}{2} F_{\mu\nu}B^{\mu\nu} - \frac{\varepsilon^{2}}{4} B_{\mu\nu}^{2} + \bar\chi (i\hat D - m_{\chi})\chi + \vert D_{\mu_{(1)}}\Phi\vert^{2} - \lambda \vert \Phi\vert^{4} + \mu_{0}^{2}\vert  \Phi\vert^{2} - b (\partial  B) + \frac{b^{2}}{2\eta},
\end{equation}
where 
%$F_{\mu\nu} = \partial_{\mu}A_{\nu} - \partial_{\nu}A_{\mu}$ with the field $A_{\mu}$ of the photon, $B_{\mu\nu} = \partial_{\mu}B_{\nu} - \partial_{\nu}B_{\mu}$ with the field %$B_{\mu}$ of DP; 
$\hat D = D_{\mu}\gamma^{\mu}$, $ D_{\mu} = \partial_{\mu} + i g B_{\mu}$; $ D_{\mu_{(1)}} = \partial_{\mu} + i g_{1} B_{\mu}$; 
$g$  and $g_{1}$ are the coupling constants (associated with the $U^{\prime} (1)$ gauge group in the dark sector)  of DP with DM and the dilaton, respectively. 
The mixing of a photon and DP is  induced by the shift $A_{\mu} \rightarrow A_{\mu} + \varepsilon B_{\mu}$;  $\lambda$ is the self-coupling constant of dilaton field, $\mu_{0}$  and $\eta$ are  real parameters. 
%The field $b(x)$ plays the role of gauge-fixing multiplier in LD (\ref{e5}), and it remains free; $\eta$ is a positive constant. The DM field $\chi$ is neutral under SM quantum numbers and charged under a hidden gauge symmetry that is broken at low energies. 
LD (\ref{e181}) is invariant under the restricted gauge transformations
$$ A_{\mu}\rightarrow A_{\mu} + \partial_{\mu}\Lambda,\,\, B_{\mu}\rightarrow B_{\mu} + \partial_{\mu}\Lambda, \,\,
 \Phi\rightarrow \Phi e^{-ig_{1}\Lambda}, \,\chi\rightarrow \chi e^{-ig\Lambda},\,\, b\rightarrow b,$$
where $\Lambda (x)$ satisfies $\Box\Lambda (x) = 0$.  
To proceed to the solution of ADDPM, we consider the real scalar fields
\begin{equation}
\label{e191}
\phi + f = \frac{1}{\sqrt 2} (\Phi + \Phi^{\star}),\,\,\, \varphi  = \frac{-i}{\sqrt 2} (\Phi - \Phi^{\star}),
\end{equation}
where 
%\begin{equation}
%\label{e20}
$ \langle\Omega,\varphi \Omega\rangle = 0, \,\,\, f = \langle\Omega, (\phi + f) \Omega\rangle. $
%\end{equation}
Having in mind  (\ref{e191}), LD (\ref{e181}) becomes:
\begin{equation}
\label{e21}
L = L_{1} + L_{2},
\end{equation}
where (the zeroth order of $g_{1}f$ and $\lambda f$ are considered)
\begin{equation}
\label{e22}
L_{1}  = -\frac{1}{4} F_{\mu\nu}^{2} - \frac {1}{2}{\varepsilon} F_{\mu\nu}B^{\mu\nu} - \frac {1}{4}{\varepsilon^{2}} B_{\mu\nu}^{2} + \bar\chi (i\partial_{\mu}\gamma^{\mu} - m_{\chi})\chi 
- g(B\cdot J) - b (\partial\cdot B) + \frac{1}{2\eta} b^{2},
\end{equation}
\begin{equation}
\label{e23}
L_{2}  =  \frac{1}{2}{m^{2}} B_{\mu}^{2} + m B_{\mu}\partial^{\mu}\varphi - \frac{1}{2}\mu^{2} \phi^{2} + 
\frac{1}{2} \left [\left (\partial_{\mu}\phi\right )^{2} + \left (\partial_{\mu}\varphi\right )^{2}\right ].
\end{equation}
In Eqs. (\ref{e22}) and (\ref{e23}), $ m = g_{1} f$ is the DP mass, $\mu = \sqrt{2}\lambda f$ is the dilaton mass. LD (\ref{e21}) thus involves six parameters: $\varepsilon$, $m_{\chi}$, $g$,  $g_{1}$, $m$ (or $f$) and  $\mu$. The mass $m$  arises through the Higgs-like mechanism. The charge of DP may be given by the measure $\rho = \varepsilon\cos\theta_{W}\,(g/e)$, where $\theta_{W}$ is the Weinberg angle, $e$ is the electromagnetic charge. Actually, $\rho\rightarrow 0$ as $g\rightarrow 0$ as well as $\varepsilon\rightarrow 0$.
The dilaton becomes massless in the limit in which conformal symmetry is recovered. The mass $\mu$ is light, proportional to the scale $f$ times the parameter $\sim \lambda$ that controls the deviation from exact scale symmetry.
  
There are four sectors in the model: SM, DM, DP and the dilaton. All the sectors connected to each other by $\varepsilon$, $g$,  $g_{1}$  or $\rho$.
The equations of motion are:
\begin{equation}
\label{e24}
m^{2} B_{\mu} + \eta\,\partial_{\mu} (\partial\cdot B) + m\,\partial_{\mu}\varphi - g J_{\mu} = 0,
\end{equation}
%\begin{equation}
%\label{e25}
$$ \Box^{2} \varphi = 0,\,\,\, \Box\varphi \neq 0. $$
%\end{equation}
The solution of Eq. (\ref{e24}) is
\begin{equation}
\label{e26}
B_{\mu} = \frac{g}{m^{2}} J_{\mu} - \frac{1}{m} \partial_{\mu}\varphi + \frac{\eta}{m^{3}} \partial_{\mu}\Box\varphi.
\end{equation}
In case of weak coupling  $g << 1$, the solution for DM field is [3]
\begin{equation}
\label{e261}
\chi (x) = \chi^{(0)} (x): \exp \{ig[1-(\eta/m^{2})\Box]\varphi (x)/m\}:,
\end{equation}
where $\chi^{(0)} (x)$ is the canonical free Dirac field of DM that commutes with $\varphi (x)$. 
%Here, 
%$  : e^{-ig\tilde\varphi} : = e^{-ig\tilde\varphi^{(+)}}  \, e^{-ig\tilde\varphi^{(-)}} , $ where 
%$\tilde\varphi (x) \simeq [-1+ (\eta/m^{2})\Box ]\varphi (x)/m $, 
%$ \tilde\varphi (x) =   \tilde\varphi^{(-)} (x) +  \tilde\varphi^{(+)} (x)$, $\tilde\varphi^{(+)} (x) = [\tilde\varphi^{(-)} (x)]^{+}$, 
%$\tilde\varphi^{(-)} (x)\Omega = 0$. 

One can easily find the inverse Higgs condition [7] enters the second term in (\ref{e26}) for DP field in ADDPM.
The two-point Wightman functions and the propagators of both DM and DP are presented in details in [3].
 
\section{Observables}
In Sec. 6 we were considered DM $\chi (x)$ and DP $B_{\mu}(x)$ as the mathematical solutions of ADDPM in the phase space $\Gamma$. Our aim is to find the solutions for DM and DP in terms of observables.
For this, we refer to L. Faddeev's approach to dynamical systems with constraints [16] where the links (constraints) $l^{a}$ ($a=1,...,k$) as the functions of canonical variables were introduced on $\Gamma$. If the canonical variables related to DM and DP do not vary throughout $\Gamma$, the following equation $l^{a} = 0$ is valid. 
The links $l^{a}$ are independent and irreducible in the sense that any arbitrary function $F$ as a linear combination of $l^{a}$, $ F = \sum_{a} c_{a}\,l^{a}$  vanishes on the physical space $M$, where $c_{a}$ are the coefficients, and being the variables, in general. The physical space $M$ is characterised by that the Poisson brackets of the link (or an arbitrary function) with themselves or with the function vanish on $M$. The latter function carries the features of the observable. In other words, the function of the interest on $M$ is an observable quantity for which the choice of arbitrary functions does not affect their variations in time. The observables are some classes of functions on $\Gamma$, or they are the functions in $M$ in the weak sense.
In our model, one can admit  the surface $ b \simeq 0$ in the physical subspace of the phase space. Neither DP $B_{\mu} (x)$ (\ref{e26}), nor DM $\chi (x)$ (\ref{e261}) are not observables because their Poisson brackets $ \{b(x), B_{\mu} (y) \} \sim \partial_{\mu} D (x-y)$ and   $\{b(x), \chi (y)\} \sim g D (x-y)$ are not equal to zero. 

Both DM and DP fields are observables only under stochastic (random) forces represented here by the operator $h_{\mu}$. DP field $B_{\mu} (x)$ has the fluctuations in medium with the probability $\Pi [B_{\mu}] \sim \exp ( - Z [h_{\mu}] )$, where 
$$ Z [ h_{\mu}] = \ln \int d B_{\mu} \exp \left \{ {\int d x [ L(x) + h_{\mu} ( x) B^{\mu} (x) ]} \right \}, $$
and $ h_{\mu}$ is a tempered distribution satisfying $\partial_{\mu} h^{\mu} (x) = \delta (x)$. $ Z[h_{\mu}]$ is entered the free energy $ F_{h}$ which is 
$$ F_{h} = \int d h_{\mu} \, Z [h^{\mu}] \exp \left \{{-\int dx h_{\mu}^{n} (x)}\right \}, $$
$ n = 1,2, ....$ are external insertions of $h_{\mu} (x)$.  The following fields for DM and DP 
\begin{equation}
\label{e177}
\tilde\chi (x;h) = \left\{\exp\left [i\,g\int d^{4} y \,h_{\mu} (x-y) \,B^{\mu} (y) \right ]\right \} \chi (x),
\end{equation}
$$\tilde B^{\mu} (x;h) = \int d^{4} y \left [ g^{\mu\nu} \,\delta (x-y) - \partial ^{\mu} h^{\nu} (x-y) \right ] B_{\nu} (y)$$
obey the following relations in Poisson brackets, respectively, $\{b(x), \tilde\chi (y;h)\} = 0,$ $ \{b(x), \tilde B_{\mu} (y;h)\} = 0, $
and, hence, $\tilde\chi$ and $\tilde B_{\mu}$ are observables. Moreover, the latter are local if $ \{O(x;h), O (y;h)\} = 0$ for all $(x-y)^{2} < 0$, where $O: \tilde\chi,\, \tilde B_{\mu}$.
%The DM field $\chi$ obeys the equation of motion: $(i\partial_{\mu}\,\gamma^{\mu} - g\,B_{\mu}\,\gamma^{\mu})\chi (x) = 0$.
In quantum case, Eq. (\ref{e6}) becomes
\begin{equation}
\label{e133}
(i\gamma^{\mu}\partial_{\mu} - m_{\chi})\tilde\chi (x;h) = g\,\gamma^{\mu}\,\tilde B_{\mu} (x;h)\,\tilde\chi (x;h). 
\end{equation}

\section{Stochastic field}
The Fourier transformation of stochastic (fluctuation) field $h_{\mu} (x)$ admits the following form:
% \begin{equation}
%\label{e14}
$$  h_{\mu} (p) = i\int d_{4} q\,q_{\mu} \left [\frac{H_{+} (p,q)}{pq + i\varepsilon} +  \frac{H_{-} (p,q)}{pq - i\varepsilon}\right ],  $$
%\end{equation}
where the positive $(+)$ and negative $(-)$ contributions are given through the form-factors  $H_{\pm}(p,q)$. The latter may also be associated with particle and anti-particle modes to stochastic field $h_{\mu} (x)$. The $q$ in $H_{\pm}(p,q)$ are internal degrees of freedom relevant to  $h_{\mu}$. The strength of the form-factor $H_{+}(p,q)$ or  $H_{-}(p,q)$ is given by the amplitude $c_{+}$ or $c_{-}$, respectively:
%\begin{equation}
%\label{e144}
$$ H_{\pm}(p,q) = c_{\pm}\,H(p,q). $$
%\end{equation}
% are the functions depending on real parameters $c_{\pm}$ in the form: $H_{+} (q) + H_{-} (p) = ( c_{+} + c_{-})H(q)$ with 
Having in mind $\partial_{\mu} h^{\mu} (x) = \delta (x)$, the form-factor is $H(p,q) = (2\,\pi)^{4}\,\delta (q-n) (c_{+} + c_{-})^{-1}$, $n$  is an arbitrary 4-vector. Thus, in $\Re_{4}$ space one has
%\begin{equation}
%\label{e1444}
$$  h_{\mu} (p) = i\, n_{\mu} \frac{1}{ (c_{+} + c_{-})} \left (\frac{c_{+}}{p\,n + i\varepsilon} +  \frac{c_{-}}{p\,n - i\varepsilon}\right ) $$
%\end{equation}
that transforms in $\Re^{4}$ to
%The latter formulae allow the following form for $ h_{\mu} (x)$:
 \begin{equation}
\label{e155}
h_{\mu} (x) = \frac{n_{\mu}}{(c_{+} + c_{-})} \int_{-\infty}^{+\infty} d\alpha \left [c_{+}\,\theta (\alpha) - c_{-}\,\theta (-\alpha)\right ]\delta (x - n\alpha). 
\end{equation}
Using (\ref{e155}) one has
 \begin{equation}
\label{e16}
\tilde\chi (x;n;c) = \exp\{ ig\int_{-\infty}^{+\infty} \frac{d\alpha}{(c_{+} + c_{-})} [c_{+} \theta (\alpha) - c_{-}\theta (-\alpha) ] n_{\mu}\,B^{\mu} (x-n\alpha)\}\chi (x),
\end{equation}
\begin{equation}
\label{e17}
\tilde B_{\mu} (x;n;c) =  B_{\mu} (x) - \partial_{\mu}\int_{-\infty}^{+\infty} \frac{d\alpha}{(c_{+} + c_{-})} [c_{+} \theta (\alpha) - c_{-}\theta (-\alpha) ] n_{\nu}\,B^{\nu} (x-n\alpha)
\end{equation}
for DM and DP observables, respectively. In case of the absence of  anti-particles ($c_{-} = 0$), taken into account  Eq. (\ref{e133}), the DM field  in $\Re_{4}$ obeys the following equation:
%\begin{equation}
%\label{e18}
$$ (\gamma^{\mu}\,p_{\mu} - m_{\chi}) {\tilde\chi} (p;n;c) =  g\,\gamma^{\mu}\int d_{4} k\,T_{\mu\nu} (k;n)\, B^{\nu} (k)\,{\tilde\chi} (p-k;n;c), $$
%\end{equation}
where $c \equiv c_{+}$ and 
%\begin{equation}
%\label{e19}
$$ T_{\mu\nu} (k;n) = g_{\mu\nu} - \frac{k_{\mu}\,n_{\nu}}{k\,n + i\varepsilon}. $$
%\end{equation}
For nonlocal DP field under the condition $n_{\mu}B^{\mu} (x - n\alpha)\rightarrow 0$, the observables for DM (\ref{e16}) and DP (\ref{e17})  become the formal solutions for these fields, $\tilde\chi (x;n;c) \rightarrow \chi (x)$ and $\tilde B_{\mu}(x;n;c) \rightarrow B_{\mu}(x)$. 

\section{Some comments to phenomenology of DP}
The phenomenology related to dark photons physics can be found  in many papers (see, e.g., [17] and the refs. therein).
The dominant DM annihilation channels are into two dilaton particles, either via exchange $\chi$ in the t-channel or directly using the operator containing two dilaton fields [18].  In the conformal QCD scenario, there is an enhanced $(33/2 - n_{light})\sim O(10)$ - factor of the effective coupling strength of the dilaton to gluons compared to that of the SM Higgs boson [19]. As a result, the main dilaton decay channel is into two gluon jets for the dilaton mass below $2 m_{W}$ [18]. However, there is the contribution from DP $\bar\gamma$ proportional to $\sim\varepsilon ^{2}$. Hence, an observation of $e^{+}e^{-}$ final pairs could unambiguously signal the discovery of new spin 1 gauge boson, DP, in the decay $\Phi\rightarrow \gamma\bar\gamma\rightarrow \gamma e^{+}e^{-}$. Using the effective couplings (\ref{e13}) and (\ref{e15}) the total decay width of the dilaton to photons and DP's is 
%\begin{equation}
%\label{e199}
$$ \Gamma_{tot} = \Gamma_{\Phi\gamma\gamma} \left [1 + {\left (\frac{\bar\alpha\,\bar {b}_{EM}}{\alpha\,b_{EM}}\right )}^{2}\,\varepsilon^{2}\right ], $$
%\end{equation}
where
$$ \Gamma_{\Phi\gamma\gamma} = \frac{\mu^{3}}{4\pi}\,g^{2}_{\Phi\gamma\gamma},\,\,\, g_{\Phi\gamma\gamma} = - \frac{\alpha\,b_{EM}}{8\,\pi\,f},$$
$\bar\alpha$ is the coupling of DP to the fermion (quark) in the loop with the mass less than the mass $\mu$ of the dilaton; $\bar{b}_{EM} \simeq b_{EM} = - 80/9$ for 
$\mu < 2 m_{W}$. In electroweak (EW) sector we use $\mu < 160\, GeV$ and $ f = 250 \,GeV$ from [18] where it has been shown a widely allowed range for a light dilaton even for $f$ not much above the EW scale. The partial decay width $\Phi\rightarrow \gamma\gamma$ is restricted by the value $\Gamma_{\Phi\gamma\gamma} < 35 \,keV$ with the coupling  $g_{\Phi\gamma\gamma} \simeq 10^{-5}\, GeV^{-1}$. The observation of an excess above the value one in the ratio
$$ R = 1 + {\left ( \frac{\bar\alpha}{\alpha}\right ) }^{2}\,\varepsilon^{2} $$
presents a challenge for the detection of new spin 1 particle with the mass $m$.
%We thus proposed to perform the sensitive search for dark photons in still unexplored area of $\varepsilon \sim 10^{-7} - 10^{-2}$ and masses of dilaton below $160\, GeV$.  The %dilaton can decay mainly to gluon jets or to two photons. If  dark photons exist, their origin is the decay of the dilaton with the sensitivity proportional to $\varepsilon^{2}$. 
The registration of dark photons is by their di-electron decay $\bar\gamma\rightarrow e^{+} e^{-}$ with the detection ratio
$$DR = R\cdot \frac{\Gamma_{\bar\gamma e^{+}e^{-}}}{\Gamma_{\bar\gamma\,all}}, $$
where in the case $m > 2 m_{e}$ [20]
$$\Gamma_{\bar\gamma e^{+}e^{-}} = \frac{1}{3}\, \alpha\,m\,\varepsilon^{2}\,{\left [ 1 - {\left  (\frac{2\,m_{e}}{m}\right )}^{2}\right ]}^{1/2}\cdot 
\left [ 1+ 2{\left (\frac{m_{e}}{m}\right )}^{2}\right ]. $$ 
We assume the decay mode $\bar\gamma\rightarrow  e^{+} e^{-}$ is dominant and, thus, $\Gamma_{\bar\gamma e^{+}e^{-}}/\Gamma_{\bar\gamma\,all} \simeq 1$.
At energies $\sim m$, the most efficient way to search for DP is through its decay to light hadrons where the decay width is $  \Gamma_{\bar\gamma \rightarrow hadrons} = \Gamma_{\bar\gamma \mu^{+}\mu^{-}}\cdot R_{had}$ with [21]
$$R_{had} = \frac{\sigma (e^{+}e^{-}\rightarrow hadrons)}{\sigma (e^{+}e^{-}\rightarrow \mu^{+}\mu^{-})} = \frac{6\pi}{\alpha}\,g_{\Phi\gamma\gamma} f,$$
and $\Gamma_{\bar\gamma \mu^{+}\mu^{-}}$ is the partial decay width of the dark photons to muon pairs. The hadron's channel indicates about $(3/4)\,b_{EM}\sim 7 $ increase of the branching ratio compared to that of the $e^{+}e^{-}$ pairs which could have profound consequences at the low energies experiments.
If no excess events are found, the obtained results can be used to impose bounds on $\varepsilon$- mixing strength as a function of DP and dilaton masses.

\section{Conclusions}
We have developed the model where DP fields fluctuate in medium. We find out that both DM and DP fields are observables only under the influence of stochastic (random) vector states. The new characteristic feature of ADDPM is that the DP field is derived by DM current and the derivatives of dilaton field. 
The inverse Higgs condition to vector field is the only natural part of general solution for DP field in ADDPM.
The observables for DM (\ref{e177}), (\ref{e16}) are defined by DP field. The latter is the natural mediator between DM and SM sectors. 
The advantage of ADDPM is that it introduces DM and DP into the theory from the beginning which otherwise only may appear after some phenomenological inputs.
We  proposed to perform the sensitive search for dark photons in still unexplored area of $\varepsilon \sim 10^{-7} - 10^{-2}$ and masses of dilaton below $160\, GeV$.  The dilaton can decay mainly to gluon jets or to two photons. If  dark photons exist, their origin is the decay of the dilaton with the sensitivity proportional to $\varepsilon^{2}$. 
%We have developed ADDPM solvable in four-dimensional space-time at lowest order of perturbative theory based on canonical quantisation. The interaction between DM and SM is %mediated by massive DP the origin of that is dictated by DM current and the derivatives of scalar dilaton fields. Because of DP fields fluctuations in medium, we find out that both DM %and DP fields are observables only under the influence of stochastic (random) vector states. The indefinite metric emerged in Green's functions does not appear  in asymptotic %expressions for physical observables.  If the dilaton is lighter than $2 m_{\chi}$, it can decay via a $\chi$-loop into a pair of Dark photons, yielding four leptons or neutrinos in the final %state. The latter can clarify a question of CP violation in dark sector through the interference of 4-lepton channel. The model considered in this paper has a phenomenological %consequence to probe DM through searching of dilatons and Dark photons in multi-lepton channels at hadron and $e^{+}e^{-}$ colliders. 

\end{document}